%% file: main_updated.tex
\documentclass[conference]{IEEEtran}
\IEEEoverridecommandlockouts

\usepackage{cite}
\usepackage{amsmath,amssymb,amsfonts}
\usepackage{algorithmic}
\usepackage{graphicx}
\usepackage{textcomp}
\usepackage{xcolor}
\usepackage{varwidth}
\usepackage{soul}
\sethlcolor{orange}

\usepackage{tikz}
\usetikzlibrary{calc}

\usepackage[T1]{fontenc}
\usepackage[english]{babel}
\usepackage{lmodern}
\usepackage{cleveref}

\usepackage{array}
\usepackage{fontawesome5}
\usepackage{tikzscale}
\usepackage{pgfplots}
\pgfplotsset{compat=1.18}
\usepgfplotslibrary{fillbetween}
\usetikzlibrary{arrows,positioning, calc}
\usetikzlibrary {arrows.meta} 

\definecolor{tabblue}{HTML}{1f77b4}
\definecolor{taborange}{HTML}{ff7f0e}
\definecolor{tabgreen}{HTML}{2ca02c}
\definecolor{tabred}{HTML}{d62728}
\definecolor{tabpurple}{HTML}{9467bd}
\definecolor{tabbrown}{HTML}{8c564b}
\definecolor{tabpink}{HTML}{e377c2}
\definecolor{tabgray}{HTML}{7f7f7f}
\definecolor{tabolive}{HTML}{bcbd22}
\definecolor{tabcyan}{HTML}{17becf}
\definecolor{deepblue}{HTML}{0000ff}

\def\BibTeX{{\rm B\kern-.05em{\sc i\kern-.025em b}\kern-.08em
    T\kern-.1667em\lower.7ex\hbox{E}\kern-.125emX}}

\usepackage{tcolorbox}
\tcbuselibrary{theorems}

\input{math_commands}

\input{macros}

\newtcbtheorem[number within=section]{defbox}{Definition}%
{colback=blue!5,colframe=blue!45!black,fonttitle=\bfseries}{def}
\newtcbtheorem[number within=section]{thmbox}{Theorem}%
{colback=red!5,colframe=red!45!black,fonttitle=\bfseries}{thm}    
\newtcbtheorem[number within=section]{codebox}{Stochastic Code}%
{colback=green!5,colframe=green!45!black,fonttitle=\bfseries}{code}    

\usepackage[final]{changes}

\begin{document}

\title{Data Compression with Stochastic Codes\\
}

\author{\IEEEauthorblockN{Gergely Flamich}
\IEEEauthorblockA{Imperial College London\\
London, UK \\
gergely.flamich@imperial.ac.uk}
\and
\IEEEauthorblockN{Deniz G{\"u}nd{\"u}z}
\IEEEauthorblockA{Imperial College London\\
London, UK \\
deniz.gunduz@imperial.ac.uk}
}

\maketitle

\begin{abstract}
Machine learning has had a major impact on data compression over the last decade 
and opened up many new theoretical and applied fields of inquiry.
\par
This paper describes one such direction -- relative entropy coding -- which focuses on constructing stochastic codes, mainly as an alternative to quantisation and entropy coding in lossy source coding.
Our primary aim is to provide a broad overview of the topic, with an emphasis on the computational and practical aspects currently missing from the literature.
\par
Our goal is threefold: for the curious reader, we aim to provide an intuitive picture of the field and convince them that relative entropy coding is a simple yet exciting emerging field in data compression research.
For a reader interested in applied research on lossy data compression, we provide an account of the most salient contemporary applications.
Finally, for the reader who has heard of relative entropy coding but has never been quite sure what it is or how the algorithms fit together, we hope to illustrate how simple and elegant the underlying constructions are.
\end{abstract}

\begin{IEEEkeywords}
relative entropy coding, channel simulation, lossy source coding, data compression, stochastic code, learned compression
\end{IEEEkeywords}

\section{From Renaissance Secret Codes to Modern Data Compression}
\label{sec:intro}
\begin{figure*}[t]
\centering
\includegraphics{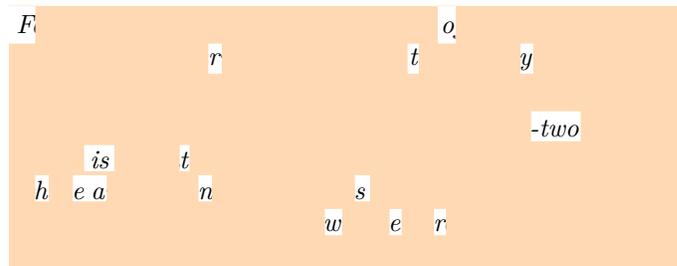}    
\caption{An illustration of a ``reverse Cardan'' grille: the cover text is fixed ahead of time, and the grille is constructed one hole at a time to reveal the next letter of the desired message.
\textbf{Left:} The first two paragraphs of ``The hitchhiker's guide to the galaxy'' by Douglas Adams \cite{adams2009hitchhiker}.
\textbf{Right:} A grille applied to the same two paragraphs, revealing an important fact about the meaning of life, the universe, and everything. }
\label{fig:reverse_cardano_grille}
\end{figure*}
\par
Writing systems emerged at least 8,000 years ago.
Soon after writing was invented, an urgent need arose that only the desired recipient be able to read the intended message, lest some ill-spirited adversary take advantage of the precious knowledge therein.
One proposed solution was steganography, the art of hiding a message in plain sight.
One of its most popular renditions was described in the middle of the 16th century by Girolamo Cardano, which is now called the \textit{Cardan grille}; though the same idea was described decades earlier by Jacopo Silvestri \cite{arnold1980apology}, and similar ideas were described centuries earlier by Wujing Zongyao in 11th-century China \cite{kahn1997codebreakers}.
\par
The Cardan grille works as follows: take a stiff piece of material, such as cardboard or metal, and cut irregularly sized holes at irregular locations; this is called the \textit{grille}.
To compose a hidden message, place the grille on top of a piece of paper and write the intended secret message in the holes.
Then lift the grille and fill the gaps between the hidden message pieces with cover text so they do not stand out.
The resulting text should appear innocuous to any unsuspecting reader, while someone with the grille can reveal the hidden message at once.
\par
Naturally, a difficulty in using the Cardan grille is that the hidden message constrains the composition of the cover message; hence, the cover text might sound awkward without proper care.
One solution is to compose the cover text first, then create an appropriate grille that reveals the desired hidden message.
We call this method the ``REverse Cardan'' (REC) method and illustrate it in \Cref{fig:reverse_cardano_grille}.
An advantage of the REC method is that the communicating parties may use any text that is available to both of them, such as the Bible or Douglas Adams' The Hitchhiker's Guide to the Galaxy.
Therefore, so long as both parties have access to the same edition of the same book, the intended recipient only needs the grille to read the hidden message.
Of course, the disadvantage now is that the grille might need to be much bigger than before: the sender composing the grille is at the whim of the pre-agreed text as to how often the pieces of the sender's intended message appear in the cover text; \Cref{fig:reverse_cardano_grille} illustrates this phenomenon well.
\par
Due to significant advances in cryptography and steganography over the last century, grilles are no longer used for serious secret communications.
However, it is interesting to study the REC method purely as a form of communication, as it underlies an emerging data compression framework called \textit{relative entropy coding}.
Thus, consider the following abstraction of the REC method: we may conceive of the shared ``cover text'' as an infinitely long string of symbols from some finite (such as the Latin) alphabet.
Furthermore, we may describe our grille as the number of symbols between consecutive holes.
For example, we can describe the first five holes of the grille in \Cref{fig:reverse_cardano_grille} by the sequence $1, 36, 37, 17, 9$.
\par
At this point, any good information theorist ought to begin pondering the question: if we represent this integer sequence using some appropriate binary code, how efficient is this protocol?
It is a beautiful fact that if 1) the symbols we wish to encode arise from a memoryless source, and 2) the symbols of the message and the shared text follow the same distribution, then we can construct a grille whose per-symbol integer description will be given by the Shannon entropy of the source.\footnote{Indeed, this result can be extended to stationary, ergodic sources using Kac's lemma.}
Of course, the assumption that the symbols of the shared text are independent and identically distributed (i.i.d.)\ is unreasonable for any book written by a human. 
However, we can easily generate the ``book'' by sampling the symbols using a pseudo-random number generator.
As such, it is customary to refer to this shared text as \textit{shared} or \textit{common randomness}.
\par
However, the real power of the REC method is that it does not require the alphabet to be finite or even discrete. Communicating parties could agree on any time-series data, such as weather forecasts, stock prices, or sensor readings.
Assuming this data is real-valued, it is futile to wait for any particular value to appear.
The remedy is to sacrifice some precision; the easiest way of achieving this is to agree that if the sender wishes to communicate some real number $X$, they send a time series index where the value of the time series at that index is, say, $\delta$ close to $X$.
Once again, it is more convenient to generate a real-valued time series using a pseudo-random number generator than to rely on external data.
\par
While these examples are cute, how is the REC method relevant to modern source coding?
In lossy source coding, given a source $X$, such as an image, the goal is to design some algorithm whose output $\hat{X}$ is ``close'' or ``similar'' to $X$ in some sense, but we need many fewer bits to encode it.
Usually, this mechanism is a function with a discrete (and sometimes finite) range, called a quantiser $\quantiser$.
This choice of mechanism constrains the conditional distribution of $\hat{X}$ to be a Dirac measure ${\Prob[\hat{X} \in A \mid X] = \Ind[\quantiser (X) \in A]}$.
As we will see, relative entropy coding, a general variant of the REC method, removes this restriction: we may choose (!) any conditional distribution $\Prob[\hat{X} \in A \mid X]$ for reconstruction.
Indeed, after giving a working definition of relative entropy coding, we dedicate the bulk of \Cref{sec:theory} to describing coding algorithms for any desired target distribution as well as highly efficient ones for special cases.
\par
A consequence of the freedom to pick any arbitrary conditional distribution for the reconstruction is that we may invert the design process for lossy compression: we choose the lossy mechanism $\Prob[\hat{X} \in A \mid X]$ that possesses whatever properties our application requires and then worry about its implementation, rather than choosing the mechanism first and then analysing its properties.
In \Cref{sec:applications}, we present three applications of this methodology:
\begin{enumerate}
\item \textbf{The ``quantisation error'' viewpoint.} 
A central challenge in quantisation is characterising and handling the quantisation error $\quantiser (X) - X$.
Relative entropy coding turns this around: we may choose distribution $\hat{X} - X$ as desired. 
An important class of examples is \textit{additive noise mechanisms}, where ${\hat{X} = X + \epsilon}$ for some random perturbation $\epsilon$ independent of $X$.
These mechanisms have significant practical relevance, as they enable \textit{learned data compression}, in which we use machine learning techniques, typically neural networks trained via gradient descent, to learn the noise mechanism.
We illustrate this via a curious nonlinear transform coding method: Bayesian implicit neural representations.
\item \textbf{Data compression with realism constraints.} 
The standard performance measure of a lossy compressor is how much it \textit{distorts} the original data, given a fixed bit budget called the \textit{rate}.
Unfortunately, with a very low bit budget, the distortion will inevitably be high.
Thus, at such low rates, a new objective emerges: to design a compression mechanism such that its outputs appear ``realistic'' \cite{hamdi2026survey}.\footnote{Indeed, this objective is meaningful at high bitrates as well, though it does not matter as much.}
As we will explain, when considering the rate-distortion-realism tradeoff, quantisers are demonstrably disadvantaged compared to arbitrary noise mechanisms that relative entropy coding can enable.
\added{Finally, on the applications side, we present an exciting line of recent work on constructing realistic, variable-rate compressors using diffusion models as a case study.}
\item \textbf{Data compression with privacy guarantees.}
A powerful framework for quantifying privacy is \textit{differential privacy} \cite{dwork2014algorithmic}, which provides a formal guarantee on how much information about sensitive user data $X$ can be inferred from released data.
To achieve differential privacy, one employs a carefully designed \textit{privacy mechanism} that randomises $X$ to obtain a privatised version $\hat{X}$.
Popular mechanisms include the additive Laplace and Gaussian mechanisms for privatising location data, such as GPS coordinates, to prevent precise tracking of an individual, and for privatising model updates in federated learning to prevent the model from memorising users' data.
As we shall explain, privacy mechanisms naturally lend themselves to relative entropy coding, yielding compression algorithms with privacy guarantees. 
%
\end{enumerate}
\par
These benefits come at a cost: general relative entropy coding algorithms are impractically slow compared with comparable quantisation-based methods. 
A central question, then, is which conditional distributions admit fast algorithms; a theme we will discuss in the next section.
\par
In summary, this article provides a broad, high-level overview of relative entropy coding and its applications.
We recommend \cite{flamich2024data} for readers interested in further computational aspects and applications to learned compression, and \cite{li2024channel} for an excellent, comprehensive exposition.
\subsection{Notation}
\par
We denote the set of integers by $\Ints$ and the set of positive integers as $\Nats$, real numbers by $\Reals$, and the set of finite-length binary strings as $\{0, 1\}^*$.
For some notation $\alpha$ and expression $\beta$, the expression $\alpha \defeq \beta$ means that $\alpha$ is defined to be $\beta$.
For integers $a, b$, we denote the set of integers between and including $a$ and $b$ as ${[a:b] \defeq [a, b] \cap \Ints}$.
For a finite set $A$, we denote the number of its elements as $\abs{A}$, for a string $s \in \{0, 1\}^*$, its length is given by $\abs{s}$.
We always round real numbers towards $+\infty$, that is $\round{x} \defeq \floor{x + 1/2}$, where $\floor{x}$ is the floor function.
Natural logarithms are denoted by $\ln$ while binary logarithms are denoted by $\lb$.
$\Ind[\cdot]$ denotes the indicator function.
For a random variable $X$ taking values in some set $\Omega$, we denote its expectation by $\Exp[X]$ and its probability distribution as $\Prob[X \in A] \defeq \Exp[\Ind[X \in A]]$ for some (measurable) set $A \subseteq \Omega$.
For two random variables $X, Y$ and a probability distribution $P$, we write $X \sim P$ to mean that $X$ has law $P$, and write $X \sim Y$ to mean that the variables are equal in distribution.
We denote uniform distributions over the interval $(a, b)$ as $\Unif(a, b)$, geometric distributions with success probability $p$ as $\Geom(p)$, exponential distributions with rate $\lambda$ as $\Exponential(\lambda)$ and normal distributions with mean $\mu$ and variance $\sigma^2$ as $\Normal(\mu, \sigma^2)$.
Furthermore, we write $X \indep Y$ to denote the independence of $X$ and $Y$.
We denote the Shannon entropy (in bits) of a discrete random variable $X$ as $\Ent[X]$.
For two probability distributions $Q$ and $P$, we denote the relative entropy/Kullback-Leibler divergence of $Q$ from $P$ as $\KLD{Q}{P}$ (in bits).
For random variables $X, Y$, we denote their mutual information as $\MI{X}{Y}$ (in bits).
Finally, $\Oh(\cdot)$ is the standard big O notation.
\section{Fantastic Stochastic Codes and How to Construct Them}
\label{sec:theory}
\par
We now give a precise
definition of stochastic and relative entropy codes; the setting is similar to source coding with its usual assumptions.
For a given input $x \in \XSpace$, our goal is to encode the output of some fixed mechanism $Y \sim P_{Y \mid X = x}$ over space $\YSpace$.
As is standard, we assume that the input/source follows some
distribution $X \sim P_X$.
The case $Y = X$ corresponds to lossless source coding, and in any other case, we think of $Y$ as a lossy representation of $X$.
Then, we look for a triplet consisting of 1) a random variable $Z$ independent of $X$, 2) an encoder function $\enc_z: \XSpace \to \{0, 1\}^*$, and 3) a decoder $\dec_z: \{0, 1\}^* \to \YSpace$. 
Observe that the encoder and decoder are actually an ensemble of functions parameterised by some variable $z$.
This triplet ``implements'' the mechanism $Y \mid X = x$: 
\begin{align}
\label{eq:rec_correctness_criterion}
\dec_Z(\enc_Z(x)) \sim P_{Y \mid X = x}
\end{align}
Note that the randomness in the above equation comes purely from the shared randomness $Z$.
From the perspective of the ``reverse Cardan method'' in \Cref{sec:intro}, $X=Y$ is the ``secret'' message we wish to encode, $Z$ is the shared ``book,'' $\enc_Z$ corresponds to constructing a grille for a given book $Z$, and $\dec_Z$ corresponds to laying the grille over the text to reveal the hidden message.
We now summarise the above discussion in the following definition.
\begin{defbox}{Stochastic Code}{stochastic_code}
For inputs $x \in \XSpace$ and conditional distributions $Y \sim P_{Y \mid X = x}$, we call the triplet $(Z, \enc_z, \dec_z)$ a \textit{stochastic code} for $Y \mid X = x$ if
\begin{enumerate}
\item $Z$ does not depend on $x$,
\item $(Z, \enc_z, \dec_z)$ satisfies \Cref{eq:rec_correctness_criterion}.
\end{enumerate}
\end{defbox}
\par
As we are interested in data compression, in addition to the ``correctness requirement'' in \Cref{eq:rec_correctness_criterion}, we also wish to minimise the expected codelength.
This desideratum results in a generalisation of Shannon's source coding theorem, which, letting $I = \MI{X}{Y}$, states that \cite{harsha2007communication,li2018strong}
\begin{align}
\label{eq:universal_relative_entropy_code_rate}
I \leq \min \Exp[\abs{\enc_Z(X)}] \leq I + 2\lb(I + 1) + \Oh(1)
\end{align}
where we minimise over stochastic codes $(Z, \enc_z, \dec_z)$.
Thus, putting everything together, in analogy to entropy codes, we define \textit{relative entropy codes}%
\footnote{%
As conditional distributions are often thought of as channels in information theory, and $(Z, \enc_z, \dec_z)$ simulates a conditional distribution, this concept is also known as \textit{channel simulation}\cite{bennett2002entanglement,li2024channel}, \textit{channel synthesis} \cite{cuff2013distributed}, \textit{coordinated sampling} \cite{cuff2010coordination}, or \textit{reverse channel code} \cite{theis2022algorithms} in the literature.
Following \cite{flamich2024data}, we distinguish these from relative entropy codes (Definition~\ref{def:relative_entropy_code}) by identifying them with stochastic codes  (Definition~\ref{def:stochastic_code}).
This distinction is analogous to lossless source codes versus entropy codes: the former requires only reconstructibility, while the latter also requires a short codelength.%
}
as
\begin{defbox}{Relative Entropy Code}{relative_entropy_code}
Given $X, Y \sim P_{X, Y}$, a relative entropy code for ${Y
\mid X}$ is a stochastic code $(Z, \enc_z, \dec_z)$ satisfying \Cref{eq:universal_relative_entropy_code_rate}.
\end{defbox}
With these definitions in hand, we will spend the rest of this section examining ways of constructing relative entropy codes; that is, how one should pick $Z$, $\enc_z$, and $\dec_z$.
Although we aimed for simplicity when possible, some parts are necessarily technical.
Readers more interested in the potential applications of these codes may take their existence for granted and skip ahead to \Cref{sec:applications}.
\begin{figure}[t]
\centering
\includegraphics[width=\linewidth]{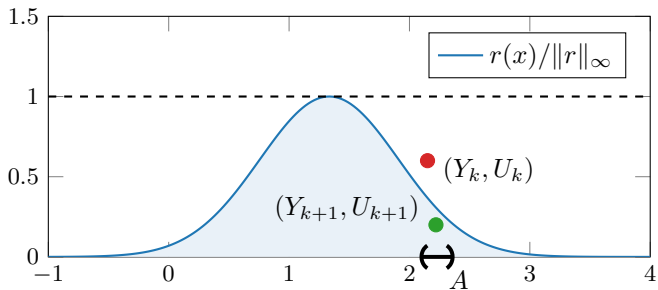}
\caption{An illustration of rejection sampling with target $Q = \Normal(1, 1/4)$ and $P = \Normal(0, 1)$.
Rejection sampling simulates points $(Y_k, U_k)$ with $Y_k \sim P$ and $U_k \sim \Unif(0, 1)$. 
If the simulated point falls within the blue-shaded region, the algorithm terminates.
The illustration also demonstrates the myopic nature of rejection sampling: two consecutive samples, $Y_k$ and $Y_{k + 1}$, might fall in the same small set $A$, but only the latter sample $Y_{k + 1}$ is accepted.
}
\label{fig:rejection_sampling}
\end{figure}
\subsection{Relative Entropy Codes from Selection Samplers}
\label{sec:selection_samplers}
\par
The decoder of any stochastic code must yield a sample from some prescribed probability distribution.
As such, it should not be surprising that computational statistics is our first port of call for inspiration.
\subsubsection{Rejection Sampling}
We begin with the simplest of samplers: the humble \textit{rejection sampler}.
The usual situation in which the rejection sampler is applied is when we wish to simulate samples from a \textit{target distribution} $Q$, but only have access to the following computational primitives:
\begin{enumerate}
\item A \textit{proposal distribution} $P$: we assume we can generate samples with law $P$.
\item The Radon-Nikodym derivative $r \defeq dQ/dP$, which we assume we can evaluate.
Readers unfamiliar with measure-theoretic mumbo-jumbo should note that when $Q$ and $P$ admit probability density functions (or mass functions in the discrete case) $q$ and $p$, then $r = q/p$.
As this will be the case for the rest of the paper, we refer to $r$ as the
\textit{density ratio} for simplicity.
Furthermore, we will always assume that $r$ is bounded from above, and that we can compute its least upper bound,%
\footnote{%
Any upper bound on $r$ is sufficient to run rejection sampling, but using the least one $\norm{r}_\infty$ will minimise the runtime and codelength.
}
which we denote as $\norm{r}_\infty$.
\end{enumerate}
Given these capabilities, rejection sampling has a simple geometric interpretation, illustrated in \Cref{fig:rejection_sampling}: it simulates a point that falls below the scaled graph of $r$.
Concretely, consider a point $(Y, U)$ with $Y \sim P$ and $U \sim \Unif(0, 1)$.
Now, a simple calculation shows that conditioned on the event that $(Y, U)$ falls under the scaled graph of $r$, the first coordinate will be $Q$-distributed:
\begin{align}
\label{eq:rejection_sampler_output_distribution}
Y \,\,\Bigg\vert\, \left\{U \leq \frac{r(Y)}{\norm{r}_\infty}\right\} \sim Q
\end{align}
\Cref{eq:rejection_sampler_output_distribution} suggests the following algorithm to simulate $Q$-distributed samples.
At each step $k \in \Nats$:
\begin{enumerate}
\item Generate $Y_k \sim P$ and $U_k \sim \Unif(0, 1)$.
\item If $U_k / r(Y_k) \leq \norm{r}_\infty^{-1}$, return $Y_k$; else go to step $k + 1$.
\end{enumerate}
If we let $K$ denote the step at which rejection sampling terminates, then \Cref{eq:rejection_sampler_output_distribution} ensures that $Y_K \sim Q$.
\subsubsection{The Basic Rejection Code}
We now construct a stochastic code $(Z, \enc_z, \dec_z)$ from rejection sampling.
Since, according to \Cref{eq:rec_correctness_criterion}, the goal is to encode a sample from $P_{Y \mid X = x}$, it stands to reason that we set the rejection sampler's target distribution ${Q \gets P_{Y \mid X = x}}$.
What proposal distribution should we pick?
We may pick essentially any distribution whose density ratio with $P_{Y \mid X = x}$ is bounded for all $x$.
However, Definition~\ref{def:relative_entropy_code} provides us with a canonical
choice: the marginal $P_Y$; thus, we set $P \gets P_Y$.
Finally, we set ${r_x(y) \defeq \frac{dP_{Y \mid X}}{dP_Y}(y \mid x)}$, which we shall assume to be bounded in $y$ for all $x$.
\par
Next, what should we choose for the common randomness $Z$?
With rejection sampling, the only candidates we could pick are the generated points $(Y_k, U_k)$.
Since, by our choice for $P$, all of them are independent of $X$, there is no harm in assuming all of them are shared.
As such, let us set $Z^{\RS} \gets \{(Y_k, U_k)\}_{k = 1}^\infty$.
There are no practical issues with this choice: if we assume the communicating parties share the same pseudo-random number generator (PRNG) and use the same seed, they can both simulate $Z^\RS$.
\par
Given what we have so far, the choice for the encoder and the decoder should also be almost clear: for the encoder, we pick the rejection sampling selection rule
\begin{align}
\label{eq:rejection_sampler_selection_rule}
K(x, Z^\RS) \defeq \min\{k \in \Nats \mid U_k / r_x(Y_k) \leq \norm{r_x}_\infty^{-1} \}
\end{align}
and for the decoder, we pick
\begin{align}
\label{eq:rejection_sampler_decoder_rule}
Y(k, Z^\RS) \defeq Y_k \quad \text{where }(Y_k, U_k) \in Z^\RS
\end{align}
We can see that $Y(K(x, Z^\RS), Z^\RS) \sim P_{Y \mid X = x}$ as required by \Cref{eq:rec_correctness_criterion}.
There remains one small problem: $K$ is an integer, not a bit string.
We can remedy the issue by using a universal code, such as the Elias $\delta$-code \cite{elias2003universal}.
While the $\delta$ code is quite simple, its precise details are not important for our present discussion; the reader only needs to know that the $\delta$ code is an invertible function $\delta: \Nats \to \{0, 1\}^*$ that encodes a positive integer $k$ using at most
\begin{align}
\label{eq:delta_code_length}
\abs{\delta(k)} \leq \lb k + 2\lb(1 + \lb k) + 1~\text{bits.}
\end{align}
Finally, we construct the rejection code as follows:
\begin{codebox}{Rejection Code}{rejection_sampling_code}
With $Y_k \sim P$, $U_k \sim \Unif(0, 1)$ and $K(x, z)$ and $Y(k, z)$ given by \Cref{eq:rejection_sampler_selection_rule,eq:rejection_sampler_decoder_rule}, respectively, the rejection code is given by:
\begin{equation*}
\begin{split}
Z^{\RS} &\defeq \{(Y_k, U_k)\}_{k = 1}^\infty \\
\enc_z^\RS(x) &\defeq \delta(K(x, z)) \\
\dec_z^\RS(s) &\defeq Y(\delta^{-1}(s), z)
\end{split}
\end{equation*}   
\end{codebox}
\subsubsection{Is our construction a relative entropy code?}
The triplet $(Z^\RS, \enc_z^\RS, \dec_z^\RS)$ is a stochastic code, but what about its expected codelength?
The first thing to note is that the samples and rejection decisions at each step of rejection sampling are independent of other steps and identically distributed.
Hence, the number of steps $K(x, Z^\RS)$ until the first acceptance is by definition a geometric random variable with mean $\norm{r_x}_\infty$.
An application of Jensen's inequality shows that $\Exp[\lb K(x, Z^\RS)] \leq \lb \norm{r_x}_\infty$.
Putting this fact together with \Cref{eq:delta_code_length} and another application of Jensen's inequality shows that the expected codelength of the rejection code is
\begin{align*}
\Exp[\abs{\enc^\RS_{Z^\RS}(X)}] &= \Exp[\lb \norm{r_X}_\infty] + \Oh(\,\lb(\Exp[\lb \norm{r_X}_\infty] + 1)\,)
\end{align*}
Thus, if ${\Exp[\lb \norm{r_X}_\infty] \approx \MI{X}{Y}}$, \Cref{eq:universal_relative_entropy_code_rate} shows that the rejection code is a relative entropy code.
This is indeed sometimes the case:
\begin{itemize}
\item If $P_{Y \mid X = x}$ is a uniform distribution for all $x$, then 
\begin{align*}
\lb\norm{r_x}_\infty = \KLD{P_{Y \mid X = x}}{P_Y}
\end{align*}
and hence 
\begin{align*}
\Exp[\lb \norm{r_X}_\infty] = \MI{X}{Y}   
\end{align*}
regardless of $P_X$.
A popular example of this is the uniform additive setting, where $Y = X + U$ with $U \sim \Unif(-1/2, 1/2)$.
However, dithered quantisation is preferable to the rejection code in this case, see \Cref{sec:dithered_quantisation}.
\item When $P_X = \Normal(0, \sigma^2)$ and $P_{Y \mid X = x} = \Normal(x, \rho^2)$, we have $\Exp[\lb \norm{r_X}_\infty] = \MI{X}{Y} + \frac{1}{2}\lb(e)$.
However, while the average codelength is technically optimal, the algorithm is of little use, since its expected runtime is infinite \cite{flamich2023adaptive,flamich2024data}.
\item If $\Exp[\lb \norm{r_X}_\infty] < C$ for some constant $C > 0$, the rejection sampler-based code is a relative entropy code according to our definition, though the constants involved might be large \cite{flamich2024data}.
\end{itemize}

Ideally, we would like a scheme for constructing stochastic codes that are always relative entropy codes.
At present, there are two known approaches: 1) to reorder the points the rejection sampler simulates, and 2) to slightly alter the rejection criterion to allow non-uniform scaling of the density ratio $r$, as opposed to scaling it uniformly with $\norm{r}_\infty^{-1}$.
The reordering approach leads to the so-called \textit{Poisson functional representation} \cite{li2018strong}, which is a special case of A* coding \cite{maddison2016poisson,flamich2022fast}. 
On the other hand, the non-uniform scaling approach yields a class of algorithms known as greedy rejection samplers \cite{harsha2007communication,flamich2023adaptive,flamich2023greedy}.
Although there is a powerful geometric intuition behind the greedy approaches (especially \cite{flamich2023greedy}), their proper construction is too technical for the present paper, and we refer the interested reader to the original material and Chapter 4 of \cite{flamich2024data}. 
We present the reordering approach below.
\subsubsection{Reordered rejection sampling and the Poisson functional representation}
\par
It is insightful to consider why the rejection code does not always yield a relative entropy code.
As we saw in its codelength analysis, the expected rate depends on the selected index $K$: the smaller $K$ is, the better the expected rate.
However, rejection sampling makes no attempt to minimise the accepted index.
To see this, consider that at step $k$, the rejection sampler proposes a sample $Y_k$ from some set $A$ and rejects it.
Now, it is perfectly possible that the next step $Y_{k + 1}$ also falls in $A$ but is accepted; see \Cref{fig:rejection_sampling} for an illustration.
Indeed, at the extreme, we could even have $r(Y_k) = r(Y_{k + 1})$, yet $Y_k$ is rejected while $Y_{k + 1}$ is accepted.
\par
This undesirable behaviour occurs because it is possible that $U_{k + 1} < U_k$.
This observation motivates the following idea: once the rejection sampler terminates at step $K$, we take the simulated random variables $(Y_1, U_1), \dotsc, (Y_K, U_K)$ and sort them in ascending order of the $U_k$s.
To this end, we define the index sorting function for common randomness $Z^\RS$, termination time $K$, and for an index $k \in [1:K]$ as follows:
\begin{align*}
\sort_{Z^\RS}(k \mid K) \,\,\defeq\,\, \abs{\{(Y_i, U_i) \in Z^\RS \mid i \leq K, U_i \leq U_k\}}
\end{align*}
What happens when we apply this function to the index $K$ of the accepted sample?
Letting $N \defeq \sort_{Z^\RS}(K \mid K)$, we see by the definition of the index sorting function that $N \leq K$, which makes $N$ a good candidate for the index to be encoded.
Moreover, since the receiver knows $Z^\RS$, they ``only'' need to know $K$ to compute $N$.
Of course, encoding $K$ just so we can send $N$ does not make sense, since that is what we wanted to avoid in the first place!
Fortunately, there are two direct fixes to this issue that were developed almost at the same time: one approach \cite{flamich2024data} is to encode $K$'s order of magnitude, $\floor{\lb K}$, and then encode $N' = \sort_{Z^\RS}(K \mid 2^{\floor{\lb K} + 1})$; the second approach \cite{phan2025channel} is to assume some additional knowledge about $K$, such as its expectation, and use this extra information to be able to encode $N$ without having to encode $K$.
Both approaches yield relative entropy codes; thus, with a bit of ``post-processing,'' we can use rejection sampling to construct efficient stochastic codes after all.
\par
However, there is a third approach, which is even more elegant: what if we could directly simulate the points $(Y_i, U_i)$ in ascending order of the $U_i$s instead of simulating them in a random order and then sorting?
This approach is precisely what the Poisson functional representation \cite{li2018strong}, also known as global-bound A* sampling \cite{maddison2014astar,maddison2016poisson,flamich2022fast}, is and does.
Unfortunately, the full exposition of this algorithm requires Poisson process theory.
Hence, we only sketch the basic arguments and refer the interested reader to \cite{li2018strong}, Chapters 4 and 5 of \cite{flamich2024data}, and Chapter 3 of \cite{li2024channel}.
\par
To reiterate, our goal is to derive a stochastic code that ``behaves'' as if we ran a rejection sampler and reordered the simulated uniform random variables.
At the heart of it all is a distributional identity involving Poisson processes.
For the purposes of this paper, we will call a sequence of random variables $(T_1, T_2, \dotsc)$ a Poisson process if the difference between any two consecutive elements is exponentially distributed with rate $1$, that is, for all $k \in \Nats$: 
\begin{align*}
T_{k} - T_{k - 1} \sim \Exponential(1)
\end{align*}
By convention, $T_0 \defeq 0$. 
%
Now, we have the following identity:
\begin{thmbox}{Ordered uniforms identity}{order_stat_identity}
Let $k \in \Nats$ and $T_{k + 1}$ be the $(k + 1)$st point of a Poisson process.
Then, conditioning on $T_{k + 1}$, the first $k$ normalised points of the process 
\begin{align*}
\left(\frac{T_1}{T_{k + 1}}, \frac{T_2}{T_{k + 1}}, \dotsc, \frac{T_k}{T_{k + 1}}\right) \,\Big\vert\,\, T_{k + 1}   
\end{align*}
are jointly equal in distribution to $k$ i.i.d. $\Unif(0, 1)$ random variables, sorted in ascending order.
\end{thmbox}
For example, when $k = 2$, for ${U_1, U_2 \sim \Unif(0, 1)}$:
\begin{align*}
(\min\{U_1, U_2\}, \max\{U_1, U_2\}) \sim \left(\frac{T_1}{T_3}, \frac{T_2}{T_3}\right) \,\,\Big\vert\,\, T_3
\end{align*}
\par
But, how can we obtain $N$ and $K$ from this?
First, to obtain the number of simulated samples $K$, we look at the acceptance criterion: rejection sampling terminates as soon as it samples a point $(Y, U)$ such that $U / r(Y) \leq \norm{r}_\infty^{-1}$.
By Theorem~\ref{thm:order_stat_identity}, $U$ must be equal to $T_j / T_{K + 1}$ for some $j$ and likewise $Y$ corresponds to some $Y_j$.
Rearranging, we see that the algorithm terminates if ${T_j / r(Y_j) \leq T_{K + 1} / \norm{r}_\infty}$.
Sadly, we do not know which $T_j / T_{K + 1}$ corresponds to $U$.
However, serendipitously, it turns out this does not matter: observe that if at some step $k$ it holds that 
\begin{align*}
\min_{j \in [1:k]}\frac{T_j}{r(Y_j)} \leq \frac{T_{k + 1}}{\norm{r}_\infty}
\end{align*}
we can be certain that there is a $j \in [1:k]$ such that for $U = T_j / T_{k + 1}$ the acceptance criterion will hold.
The beauty of this observation is that we can check the above criterion sequentially: letting $\tau_k = \min_{j \in [1:k]}T_j/r(Y_j)$ be the running minimum at step $k$, the algorithm stops when $\tau_k \leq T_{k + 1}/ \norm{r}_\infty$.
Thus, we define 
\begin{align*}
K \defeq \min\left\{k \in \Nats \,\,\middle\vert\,\, \tau_k \leq \frac{T_{k + 1}}{\norm{r}_\infty} \right\}
\end{align*}
We emphasise that since all we are doing under the hood is ``simulating rejection sampling in $U_k$ order,'' the $K$ above and the runtime of standard rejection sampling are equal in distribution.
We use a similar technique to find which point satisfies the acceptance
criterion.
Letting ${N(k) = \argmin_{j \in [1:k]}\{\,T_j/r(Y_j)}\,\}$, the returned index is $N = N(K)$.
Now, although we redefined the algorithm using a Poisson process, our argument above shows that 
\begin{align*}
N \sim \sort_{Z^\RS}(K \mid K)  
\end{align*}
To summarise, the Poisson functional representation / A* sampling works as follows.
Set $T_0 \gets 0, \tau \gets \infty$ and $N \gets 0$.
Then, at each step $k$:
\begin{itemize}
\item Simulate $Y_k \sim P$ and $\Delta_k \sim \Exponential(1)$.
\item Put $T_k \gets T_{k - 1} + \Delta_k$.
\item If $\tau < T_k / \norm{r}_\infty$, terminate and return $Y_{N}, N$.
\item If $T_k / r(Y_k) < \tau$, set $\tau \gets T_k/r(Y_k)$ and $N \gets k$.
\item Go to step $k + 1$
\end{itemize}
We can turn this sampling algorithm into a stochastic code analogously to how we constructed the code from the original rejection sampler.
For input $x$, set the target distribution to $Q \gets P_{Y \mid X = x}$, the proposal to $P \gets P_Y$ and ${r_x(y) \defeq \frac{dP_{Y \mid X}}{dP_Y}(y \mid x)}$.
Now, put $Z^\AS \gets \{(Y_n, T_n)\}_{n = 1}^\infty$ for the common randomness and define
\begin{align}
K(x, Z^\AS) &\defeq \min\left\{k \in \Nats \,\,\middle\vert\,\, \min_{j \in [1:k]}\left\{\frac{T_j}{r_x(Y_j)}\right\} < \frac{T_{k + 1}}{\norm{r_x}_\infty}\right\}\nonumber\\
\label{eq:a_star_sampler_selection_rule}
N(x, Z^\AS) &\defeq \argmin_{n \in [1:K(x, Z^\AS)]}\left\{\frac{T_n}{r_x(Y_n)}\right\}\\
Y(n, Z^\AS) &\defeq Y_n, \quad \text{where } (Y_n, T_n) \in Z^\AS\nonumber
\end{align}
Then, using the Elias $\delta$-code as before, we obtain the A* code / Poisson functional representation triplet.
\begin{codebox}{A* Code}{a_star_code}
Letting $Y_n \sim P$, $\{T_n\}_{n = 1}^\infty$ be a Poisson process and setting $N(x, z)$ and $Y(n, z)$ as in \Cref{eq:a_star_sampler_selection_rule}, the A* code is given by
\begin{equation*}
\begin{split}
Z^\AS &\defeq \{(Y_n, T_n)\}_{n = 1}^\infty \\
\enc_z^\AS(x) &\defeq \delta(N(x, z)) \\
\dec_z^\AS(s) &\defeq Y(\delta^{-1}(s), z)
\end{split}
\end{equation*}   
\end{codebox}
It is a beautiful theorem of Li and El Gamal \cite{li2018strong} that the A* code above is always a relative entropy code.
\subsubsection{Selection Samplers}
\begin{figure}[t]
\centering
\includegraphics[width=\linewidth]{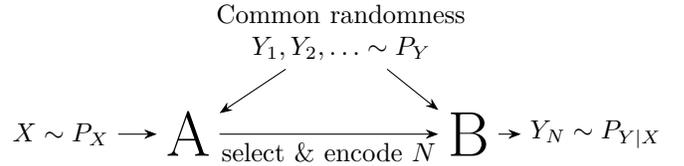}
\caption[Illustration of relative entropy coding using a selection sampler.]{Illustration of relative entropy coding for the pair of random variables $X, Y \sim P_{X, Y}$ using a selection sampler.
The sender \textbf{A}nna and the receiver \textbf{B}{\'e}la share a sequence of i.i.d.\ $P_Y$-distributed samples as their common randomness $Z$.
Then, upon receiving a source sample $X \sim P_{X}$, \textbf{A} uses a selection rule $N$ that selects one of the samples in the shared sequence such that $Y_N \sim P_{Y \mid X}$.
Since the selected index $N$ is discrete, \textbf{A} uses the Elias $\delta$-code to efficiently encode $N$ and transmit it to \textbf{B}.
Finally, \textbf{B} can recover a $P_{Y \mid X}$-distributed sample by decoding $N$ and selecting the $N$th sample in the shared sequence.
}
\label{fig:selection_sampling_sketch}
\end{figure}

It is insightful to take stock of the pattern of constructing general-purpose stochastic codes.
To this end, we consider selection samplers \cite{flamich2024data}:
\begin{defbox}{Selection Sampler}{selection_sampler}
Let $Q$ be a probability distribution and $(Y_1, Y_2, \dotsc)$ an infinite sequence of i.i.d.\ samples, where $Y_i \sim P$.
We refer to $Q$ as the target distribution and to $P$ as the proposal distribution.
Then, a \textit{selection sampler} for $Q$ using the sequence $(Y_1, Y_2, \dotsc)$ is defined by two random variables, $N$ and $K$, called the \textit{selection rule} and the \textit{runtime}, respectively, taking values in $\Nats$ and obeying the following properties:
\begin{enumerate}
\item The sample picked by the selection rule $N$ is always $Q$-distributed: $Y_N \sim Q$.
\item The runtime is independent of the future: the event $\{K \geq k\}$ that the algorithm does not terminate at step $k$ is independent of all future samples $(Y_k, Y_{k + 1},\dotsc)$.
\item The selection rule always picks a sample that we already examined: $1 \leq N \leq K$.
\end{enumerate}   
\end{defbox}
We see that both the rejection sampler and the Poisson functional representation are selection samplers.
Given a selection sampler, we can follow the recipe we used before to construct a stochastic code: given an input $x$, set the target distribution as $Q \gets P_{Y \mid X = x}$, and the proposal as $P \gets P_Y$.
Then, we set a sequence of i.i.d.\ samples from $P_Y$ as the common randomness $Z^{\SS} \gets \{Y_i\}_{i = 1}^\infty$, where $Y_i \sim P_Y$.
Next, let $N(x, Z^{\SS})$ and $K(x, Z^{\SS})$ be the selection rule and runtime of the sampler for $P_{Y \mid X = x}$ using the proposal sequence $Z^\SS$, respectively.
Finally, let $Y(n, Z^{\SS})$ denote the $n$th element of $Z^\SS$.
Then, we obtain the selection code; see \Cref{fig:selection_sampling_sketch} for an illustration.
\begin{codebox}{Selection Code}{selection_sampler_code}
Let $X, Y \sim P_{X, Y}$.
Now, let $\{Y_i\}_{i = 1}^\infty$ with $Y_i \sim P_Y$, and let $N(x, z)$ and $K(x, z)$ be a selection sampler for $P_{Y \mid X = x}$.
Then, the selection code is given by
\begin{equation*}
\begin{split}
Z^\SS &\defeq \{Y_i\}_{i = 1}^\infty \\
\enc_z^\SS(x) &\defeq \delta(N(x, z)) \\
\dec_z^\SS(n) &\defeq Y(\delta^{-1}(n), z)
\end{split}
\end{equation*}   
\end{codebox}
To demonstrate the utility of the selection sampler abstraction, we turn to an important practical aspect of relative entropy coding algorithms: their runtime.
As we saw, the runtime of both rejection sampling and the Poisson functional representation given input $X = x$ is geometrically distributed with mean $\norm{r_x}_\infty$.
Hence, using the fact that ${\Exp[\lb r_x(Y)] \leq \lb \norm{r_x}_\infty}$ and Jensen's inequality, we find that
\begin{align}
\label{eq:general_rec_exponential_runtime}
2^{\MI{X}{Y}} \leq \Exp[\norm{r_X}_\infty]
\end{align}
\Cref{eq:general_rec_exponential_runtime} is bad news: it shows that the average number of samples the algorithm needs to generate scales at least (!) exponentially in the information content.
For comparison, entropy coding algorithms such as arithmetic coding scale linearly in the information content.
\par
It is thus natural to wonder whether we could find a selection sampler with a faster average runtime.
Unfortunately, this is not so, as Theorem 3.2.1 of \cite{flamich2024data} shows: \textbf{the expected runtime of any selection sampler is at least $\Exp[\norm{r_X}_\infty]$}. 
This result means that the runtime of the Poisson functional representation is optimal.
Readers familiar with computational statistics should not find the above result surprising: sampling is generally a computationally hard problem.
One may also view selection sampling-based relative entropy coding as randomised vector quantisation.
This view provides some intuition for the necessarily slow runtime: general vector quantisation is known to be a hard computational problem.
\par
However, not all hope is lost: the theorem merely precludes the existence of a universal, fast relative entropy coding algorithm.
When $P_{X, Y}$ admits some structure, it is in some cases possible to utilise it to construct fast relative entropy coding algorithms, as we shall do next. 
%
%
\subsection{Relative Entropy Codes from Dithered Quantisers}
\label{sec:dithered_quantisation}
\par
We now take the opposite approach to the previous section: we develop a relative entropy code for additive uniform channels of the form $Y = X + U$ only, where $U \sim \Unif(-1/2, 1/2)$.
Though this might first appear to be a significant restriction, we will use it as a stepping stone to construct codes for a wider variety of channels.
\par
The basic building block is the following identity:
\begin{thmbox}{Dithering Identity \cite{zamir2014lattice}}{dithering_identity}
Let $x \in \Reals$ and $U, U' \sim \Unif(-1/2, 1/2)$.
Then:
\begin{align*}
\round{x + U} - U \,\,\sim\,\, x + U'
\end{align*}
\end{thmbox}
Since $N = \round{x + U}$ is discrete, the dithering identity suggests a straightforward way to construct a relative entropy code for additive uniform channels.
We set the common randomness as $Z^\DQ \gets U$ and $N(x, u) = \round{x + u}$.
Similar to the stochastic codes of the previous section, the final step is to turn $N(x, Z^\DQ)$ into bits using some code.
While we could use the $\delta$-code again, it turns out we can use something much better in this case:
we can construct an entropy code $C_u: \Ints \to \{0, 1\}^*$, such as a Huffman code, to encode $N(x, u)$.
As with the $\delta$-code, the precise details of entropy codes are not important for the present discussion; the reader only needs to know the following. 
First, to construct an entropy code $C_u$ for $N(X, u)$, we need to be able to compute its distribution.
Second, the expected codelength of an entropy code $C_u$ for $N(X, u)$ is, well, the Shannon entropy $\Ent[N(X, U)]$:
\begin{align*}
\Exp[\abs{C_U(X)}] = \Ent[N(X, U)] + \Oh(1)
\end{align*}
Fortunately, in the case of dithered quantisation, we can indeed work out the distribution of $N(X, u)$ by expressing it in terms of the distribution of $X$ (see Chapter 5 of \cite{zamir2014lattice}): 
\begin{align*}
\Prob[N \leq n \mid U = u] 
= \Prob\left[X \leq n - u + \frac{1}{2}\right] 
\end{align*}
Thus, let $C_{u}(n)$ be some entropy code (such as a Huffman code) using the above distribution for $N$.
Then, we can construct the dithered quantisation-based stochastic code:
\begin{codebox}{Dithered Quantiser}{dithered_quantisation_code}
Let $X \sim P_X$ and for $U \sim \Unif(-1/2,  1/2)$, let $Y \defeq X + U$.
Letting $C_u$ be the entropy code given above, the dithered quantiser code is
\begin{equation*}
\begin{split}
Z^\DQ &\defeq U  \\    
\enc_u^\DQ(x) &\defeq C_u(\round{x + u}) \\
\dec_u^\DQ(n) &\defeq C_u^{-1}(n) - u
\end{split} 
\end{equation*}   
\end{codebox}
This code turns out to be very efficient: it can be shown (see, for example, 5.2.1 of \cite{zamir2014lattice}) that
\begin{align}
\label{eq:dithered_quantiser_rate}
\Exp[\abs{\enc_{Z^\DQ}^\DQ(X)}] = \MI{X}{Y} + \Oh(1)
\end{align}
Furthermore, a significant advantage of dithered quantisation is that it is blazingly fast: computing $N(x, u)$ involves only a few basic arithmetic operations, and the average time it takes to encode it with any one of the common entropy codes scales linearly in $\Ent[N(X, U)]$!
\par
\subsubsection{Channels are like onions: layered quantisation}
\label{sec:layered_quantisation}
The fact that dithered quantisation is only applicable to additive uniform perturbations might seem quite limiting at first.
However, Hegazy and Li \cite{hegazy2022randomized} showed that we can use dithered quantisation to construct a relative entropy code for any one-dimensional additive unimodal mechanism. 
The main idea is to find a \textit{scale mixture of shifted uniforms} (SMSU) representation for the additive noise: 
\begin{defbox}{Scale Mixture of Shifted Uniforms Representation}{smsu}
Let $\epsilon$ be a real-valued random variable with a unimodal probability density function.
An SMSU representation for $\epsilon$ is the triplet $(U, S, b)$, where $U \sim \Unif(-1/2, 1/2)$, $S$ is a positive random variable and $b(s)$ is a function such that
\begin{align}
\label{eq:scale_mixture_of_unifs}
\epsilon \sim S\cdot (U + b(S))
\end{align}
\end{defbox}
It can be shown that the conditions on $\epsilon$ in Definition~\ref{def:smsu} ensure that $\epsilon$ always admits an SMSU representation.
Taking Gaussians as an example, one can show that letting the scale be twice a $\chi$-distributed variable with 3 degrees of freedom $S \sim 2 \cdot \chi(3)$ and setting $b(s) = 0$, we have $S \cdot U \sim \Normal(0, 1)$.
\par
To construct a relative entropy code, we combine the above with the dithering identity: for any $x \in \Reals$, and $U, U' \sim \Unif(-1/2, 1/2)$ conditioned on $S$, we have
\begin{align*}
\round*{\frac{x}{S} + b(S)+ U} \!- U\,\, \sim \,\,\frac{x}{S} + b(S) + U' \tag{Theorem~\ref{thm:dithering_identity}}
\end{align*}
Multiplying both sides by $S$, \Cref{eq:scale_mixture_of_unifs} guarantees
\begin{align}
\label{eq:layered_quantiser_identity}
S \left(\round*{\frac{x}{S} + b(S)+ U} - U \right) \sim x + \epsilon
\end{align}
From a quantisation perspective, \Cref{eq:layered_quantiser_identity} corresponds to randomly scaling and shifting the quantisation grid to achieve the desired distribution.
This way, we can also extend the dithered quantiser code to any one-dimensional, unimodal additive noise mechanism $\epsilon$.
First, we set the common randomness to $Z^\LQ \gets (U, S)$, where $(U, S)$ are the pair of random variables from the SMSU representation in \Cref{eq:scale_mixture_of_unifs}.
As before, we are interested in encoding ${N \defeq \round*{\frac{x}{S} + b(S)+ U}}$.
As with the original dithered quantiser, we can express the distribution of $N$ in terms of the distribution of $X$:%
\begin{align*}
\Prob[N \leq n \mid U = u, S= s] =\Prob\left[X \leq s \left(n - b(s) - u + \frac{1}{2}\right)\right]
\end{align*}
Now, let $C_{u, s}(n)$ be some entropy code over the integers, such as a Huffman code, constructed using the above distribution. 
Then, we can construct the following layered dithered quantiser-based stochastic code:
\begin{codebox}{Layered Quantiser}{layered_quantiser_code}
Let $X \sim P_X$ and let $Y \defeq X + \epsilon$ for some perturbation $\epsilon \indep X$ such that $\epsilon$ admits an SMSU representation $(U, S, b)$, as given by Definition~\ref{def:smsu}.
Finally, letting $C_{u, s}$ be the entropy code as above, the layered quantiser code is
\begin{equation*}
\begin{split}
Z^\LQ &\defeq (U, S) \\
\enc_{u, s}^\LQ(x) &\defeq C_{u, s}\left(\round*{\frac{x}{s} + b(s)+ u}\right) \\
\dec_{u, s}^\LQ(n) &\defeq s \cdot (C_{u, s}^{-1}(n) - u) 
\end{split}
\end{equation*}   
\end{codebox}
Arguments analogous to the ones leading to \Cref{eq:dithered_quantiser_rate} show that
\begin{align*}
\Exp[\abs{\enc_{Z^\LQ}^\LQ(X)}] = \MI{X}{Y \mid S} + \Oh(1)
\end{align*}
While the above is a clean expression, it is not immediately clear that it qualifies the layered quantiser code as a relative entropy code, since $\MI{X}{Y} \leq \MI{X}{Y \mid S}$.
However, using ideas from \cite{hegazy2022randomized}, it is possible to show that 
\begin{align*}
\MI{X}{Y \mid S} \leq \MI{X}{Y} + \lb(\MI{X}{Y} + 1) + \Oh(1)
\end{align*}
which shows that it is indeed a relative entropy code.
Furthermore, like dithered quantiser codes, layered quantiser codes are also optimally fast.
Thus, layered quantisers provide a complete solution for one-dimensional unimodal additive noise distributions.
However, to apply them, we must be able to derive the SMSU representation for the perturbation, which might be difficult in practice.  
\subsection{Advanced Aspects}
\par
We now touch on a few more advanced features of relative entropy codes and outline some more sophisticated constructions, which lie beyond the scope of this paper.
\subsubsection{Considerations regarding common randomness}
First, note that we put no restriction on the common randomness other than that it be independent of the input.
An important question is whether common randomness is necessary at all, and if so, how much is required.
On the first point: yes, it is necessary.
It is possible to construct stochastic codes without common randomness; however, such codes necessarily have significantly higher rates than relative entropy codes.
Several works have analysed the amount of common randomness required to construct relative entropy codes. 
However, the practical relevance of limiting the amount of common randomness remains an open question.
Essentially, our experience has been that once we share a pseudo-random number generator seed, generating as many samples as needed has no noticeable impact on the data compressor's performance.
Hence, in practice, we usually treat the cost of shared randomness as a small, constant number of bits that the encoder must use to share its random number generator seed.
We refer readers interested in a more in-depth analysis of common randomness to the monograph \cite{li2024channel}, which provides an excellent survey on the topic.
\subsubsection{Fixed-length stochastic codes}
We have discussed only variable-length relative entropy codes and made no mention of fixed-length codes, i.e., those in which every codeword has the same number of bits.
For finite, discrete channels, it is possible to construct fixed-length stochastic codes.
However, once again, the rate of such codes will be much higher than that of a relative entropy code \cite{li2024channel}.
Alternatively, we may require only that the code's output distribution be approximately $P_{Y \mid X = x}$, as measured by an appropriate statistical distance, such as total variation. We discuss this in the upcoming \Cref{sec:approximate_stochastic_codes}.
\subsubsection{Multidimensional layered quantisers}
As we saw in \Cref{sec:layered_quantisation}, layered quantisation essentially solves the relative entropy coding problem for one-dimensional, unimodal distributions almost as conclusively as one can hope for: layered quantiser codes are fast and achieve optimal codelengths.
However, while we can apply dithered quantisation to multiple dimensions without any loss of efficiency, extending any layered quantiser is highly non-trivial and, in most cases, appears hopeless.
At the heart of the matter is that the dithering identity and the SMSU representation generalise ``differently'' to higher dimensions.
Concretely, the dithering identity in higher dimensions generalises using uniform distributions over very special sets arising from lattice theory, called \textit{Voronoi cells} (of a lattice).
On the other hand, the SMSU identity extends to higher dimensions through the superlevel sets of the density function of the additive perturbation.
Thus, layered quantisation generalises straightforwardly only to cases where the superlevel sets of the additive perturbation are the Voronoi cells of a lattice.
Unfortunately, aside from the two-dimensional Laplace distribution, we are not aware of any well-known multivariate distribution in which this coincidence occurs.
An interesting approach specifically targeted at multivariate Gaussian relative entropy coding was investigated by Kobus et al.~\cite{kobus2024gaussian}, who proposed randomly rotating the Voronoi cells to make the additive noise spherical symmetric.
While their approach has a better rate than applying Gaussian layered quantisation dimensionwise, it only produces approximate samples.
\subsubsection{Sampling-as-search and branch-and-bound samplers}
In 2016, Chris Maddison observed that one could use the theory of Poisson processes to systematically construct selection samplers, such as rejection sampling and A* sampling \cite {maddison2016poisson}.
\cite{flamich2024data} extended his observation by introducing two new selection samplers based on Poisson processes.
Basically, Maddison observed that random variate simulation can be recast as a search problem over a Poisson process, leading to the \textit{sampling-as-search} paradigm.
\par
From this lens, the samplers in \Cref{sec:selection_samplers} correspond to the most basic search algorithm: \textit{linear search}.
However, the advantage of the sampling-as-search paradigm is that, when the problem at hand has a special structure, we have the vast literature on search problems at our disposal to develop more efficient sampling algorithms.
\par
One powerful idea that carries over is branch-and-bound (BnB) search, an instance of the divide-and-conquer paradigm.
At a high level, BnB search recursively partitions the search space and quickly eliminates regions that are provably not containing the object we are looking for. 
Lifting BnB search to the sampling-as-search domain gives rise to \textit{branch-and-bound samplers}.
Akin to selection samplers, we may also use BnB samplers to construct relative entropy codes.
Furthermore, it can be shown that for one-dimensional, unimodal distributions, there exist BnB relative entropy codes, such as a BnB variant of A* coding, whose expected runtime scales as $\Exp[\lb \norm{r_X}_\infty]$, an exponential improvement over selection samplers; see Chapter 5 of \cite{flamich2024data}. 
As such, BnB A* coding provides an essentially optimal alternative to layered quantisation for one-dimensional, unimodal distributions at the cost of a small, constant increase of the average codelength.
However, unlike layered quantisers, BnB A* coding is much more widely applicable: it only requires that we can evaluate the perturbation's distribution function, and we do not need to compute an SMSU representation.
\subsubsection{The gap between the minimum and achievable rate}
\added{%
Observe that our rate requirement for a relative entropy code in \Cref{eq:universal_relative_entropy_code_rate} includes a $\log(\MI{X}{Y} + 1)$ term.
This condition is unlike entropy coding, where we can always construct a code whose expected rate is within a constant of the optimal lower bound for any source. 
We also saw that, apart from the dithered quantiser, the rate of all relative entropy codes \textit{did} include a logarithmic term.%
}%
\par
\added{%
Thus, a natural question is whether the logarithmic gap between the lower and upper bounds in \Cref{eq:universal_relative_entropy_code_rate} is fundamental or whether it is reducible.
Unfortunately, it is fundamental: there are joint distributions $P_{X, Y}$ for which the optimal relative entropy code for the channel $P_{Y \mid X}$ requires at least ${\MI{X}{Y} + \log(\MI{X}{Y} + 1) - 1}$ bits \cite{li2018strong}.
There is an intuitive reason for the necessity of the log term: the ``description length'' for a sample $Y \sim P_{Y \mid X}$ is on average $\MI{X}{Y}$ bits long, but since the code is stochastic, we also need to encode the codelength for the decoder to know when to stop reading the bitstream.
More formally, for each realisation of the common randomness, we require that the stochastic code is prefix-free.
Thus, the number of bits beyond $\MI{X}{Y}$ hinges on \textit{how predictable the codelength is}.
Then, the $\log(\MI{X}{Y} + 1)$ term in the achievable rate corresponds to making no assumptions about the problem structure, or equivalently, assuming that the codelength is unpredictable.
However, for special classes of channels, we can do better.
}%
\par
\added{%
We have already seen that for lossless source coding, i.e. when $Y = X$, and for additive uniform dithering, the codelength is perfectly predictable, and hence there is no additional overhead.
Another special case are i.i.d.\ channels, where we fix some joint distribution $P_{X, Y}$ and for some $n \in \Nats$ we consider the encoding $n$ i.i.d. copies of $Y^n$ given $X^n$, that is, we aim to encode $Y^n \sim P_{Y^n \mid X^n}$.
In this case, the codelength becomes more predictable as $n$ gets large and we only need roughly $\frac{1}{2}\log(\MI{X}{Y} + 1)$ bits to describe the codelength, and in some cases the lower bound can also be tightened to include this term \cite{sriramu2024optimal,flamich2025redundancy}.%
}%
\par
\added{%
At present, the necessity of the log term appears to lead to a tradeoff between rate and computational complexity relative entropy codes in practice.
Indeed, when $\MI{X}{Y}$ is large, we might be tempted to break up the problem into smaller chunks, for example, dimensionwise when $Y$ is a vector.
However, encoding each chunk separately introduces an additional logarithmic overhead in the codelength.
Hence, the rate of the divide-and-conquer approach is strictly worse than encoding $Y \mid X$ directly.%
}%
\par
\added{%
Finally, one might take a more radical approach and reject $\MI{X}{Y}$ as the measure for characterising the rate of relative entropy codes and look for other information measures that might yield a tighter gap.
A promising candidate for this is the \textit{channel simulation divergence} \cite{goc2024channel}, which was recently shown to lead to a tighter lower bound on the rate of relative entropy codes than the mutual information \cite{flamich2024data,flamich2025redundancy}, but it is unknown if it is achievable.%
}%
\subsubsection{Approximate stochastic codes}
\label{sec:approximate_stochastic_codes}
\par
\added{%
We discussed only exact relative entropy codes so far: we required that \Cref{eq:rec_correctness_criterion} be satisfied exactly.
Unfortunately, as Theorem 3.2.1 of \cite{flamich2024data} shows, the expected runtime of general relative entropy codes scales as $\Exp_X[\norm{r_X}_\infty]$, which is strictly worse than $\exp(\MI{X}{Y})$. 
Furthermore, their runtime is also random and can fluctuate significantly in practice.
For some channels, faster algorithms exist, such as dithered and layered quantisers, so runtime is not an issue. 
However, there is a more general strategy: to sacrifice exactness in favour of a better runtime.%
}%
\par
\added{%
Indeed, we can easily turn any selection sampler $(N, K)$ (see Definition \ref{def:selection_sampler}) into an approximate algorithm via \textit{step-limitation}.
Concretely, we first fix a maximum runtime budget $B$.
Then, if the runtime of the sampler $K$ is less than $B$, we return the sampler's output.
Otherwise, if $K$ exceeds $B$, the sampler ``early-terminates'' and outputs an arbitrary value, such as a random proposal sample.%
}%
\par
\added{%
A common way to analyse such algorithms is to quantify the budget $B$ required so that the total variation error between the approximate sampler's output distribution $Q_B$ and the true target distribution $Q$ is at most $\epsilon > 0$.
Relative entropy code-based approximate samplers turn out to be very efficient: as shown in \cite{flamich2024some}, for a given target distribution $Q$ and proposal $P$, setting $B = 2^{(\KLD{Q}{P} + 1) / \epsilon}$ already guarantees that $\TVD{Q}{Q_B} \leq \epsilon$, achieving the best possible lower bound derived by Block and Polyanskiy within a small multiplicative factor \cite{block2023sample}.
}%
\par
\added{%
There are also general approximate relative entropy codes that are not constructed from step-limitation, such as \textit{minimal random coding} \cite{havasi2018minimal} and \textit{ordered random coding} \cite{theis2022algorithms}.
These find use in special-purpose applications, such as compressing differentially private mechanisms or lossy source coding with side information \cite{phan2024importance}.%
}%
\par
\added{%
Finally, two further benefits of the approximate samplers we outline above are that 1) they extend the applicability of relative entropy coding to cases where the target-proposal density ratio is unbounded, such as in the case of diffusion model-based data compression (\Cref{sec:diffc}) and 2) it also makes relative entropy coding significantly more parallelisable, as was exploited in \cite{vonderfecht2025lossy}.
}%
\subsubsection{Relative entropy codes from channel codes}
Finally, we briefly mention the recent work of Sriramu et al. \cite{sriramu2024fast}.
They start from the well-known duality between source coding and channel coding and ask a natural question: whether channel codes can also be used to construct relative entropy codes.
They answer this question positively: they use polar codes to construct relative entropy codes for high-dimensional i.i.d. Bernoulli channels, with essentially optimal runtime.
Thus, their pioneering work represents an exciting avenue for future research. 
\section{The proof of the pudding is in the applications}
\label{sec:applications}
As we have already hinted, relative entropy is best thought of as an alternative to quantisation for lossy source coding.
But what concrete benefits does it offer compared to quantisation?
In this section, we collect some of its most salient applications to date.
\subsection{Learned lossy data compression}
\label{sec:learned_data_compression}
In lossy data compression, we are given some data $X \sim P_X$ and a distance measure $d$, called the \textit{distortion}.
The goal is to design a \textit{lossy compression mechanism} that outputs some $\hat{X}$ that is cheap to encode and minimises $d(X, \hat{X})$ at the same time.
\par
We can employ relative entropy coding to obtain such a mechanism: we design a conditional distribution $P_{\hat{X} \mid X}$ such that encoding a sample $\hat{X} \sim P_{\hat{X} \mid X}$ has low distortion.
For example, we can require that on average the distortion does not exceed some pre-specified level $D$, that is, $\Exp_{X}[\Exp_{\hat{X} \mid X}[d(X, \hat{X})]] \leq D$.
What is the best we can do under this setup?
For a fixed source distribution $P_X$ and conditional distribution $P_{\hat{X}\mid X}$, relative entropy coding requires approximately $\MI{X}{\hat{X}}$ bits to encode a sample.
Thus, the best we can do is
\begin{equation}
\label{eq:source_coding_rate_distortion_function}
\begin{split}
R(D) &= \inf_{P_{\hat{X} \mid X} \in \DistFamily} \MI{X}{\hat{X}} \\
&\text{subject to}\quad \Exp_{X}[\Exp_{\hat{X} \mid X}[d(X, \hat{X})]] \leq D.
\end{split}
\end{equation}
When $\DistFamily$ contains all possible conditional distributions, $R(D)$ is known as the \textit{rate-distortion function}.
Note also that $R(D)$ is closely related to, but not equivalent to, the information rate-distortion function from classical information theory.
The two differences are that in learned compression, 1) we do not assume that the joint distribution factorises dimensionwise as $P_{\hat{X}, X} = P_{\hat{X}_1, X_1}^{\otimes \dim(X)}$ in the definition of $R(D)$, and as such 2) we do not (and cannot) analyse the asymptotic behaviour as $\dim(X) \to \infty$.%
\par
While the connection to the rate-distortion function is a nice theoretical property, relative entropy coding also offers a practical benefit: we can directly optimise the objective in \Cref{eq:source_coding_rate_distortion_function}; this is called ``end-to-end training'' in machine learning parlance.
Concretely, let $\DistFamily$ be a family of \textit{parametric, reparameterisable distributions}: we assume that for every distribution $P_{\hat{X} \mid X} \in \mathcal{D}$, we can write $\hat{X} = \phi(\theta, X, \epsilon)$ for a deterministic function $\phi$ with parameters $\theta$ and independent noise $\epsilon \perp X$.
Furthermore, we assume that $\phi$ is differentiable in its argument $\theta$; by far the most popular example of this are the additive Gaussian channels $P_{\hat{X} \mid X = x} = \Normal(\theta \cdot x, I)$, which admit the representation $\hat{X} = \theta \cdot x + \epsilon$, where $\epsilon \sim \Normal(0, I)$.
Now, borrowing a standard trick from optimisation, we write down the Lagrange dual of the objective in \Cref{eq:source_coding_rate_distortion_function}:
\begin{align}
\label{eq:source_coding_rate_distortion_objective}
\mathcal{L}(\theta, \beta) = \MI{X}{\phi(\theta, X, \epsilon)} + \beta \cdot \Exp[d(X, \phi(\theta, X, \epsilon))] + \Oh(1)
\end{align}
The relevance of the Lagrange dual is that a minimiser of $\Loss(\theta, \beta)$ for each $\beta$ corresponds to $R(D(\beta))$ from \Cref{eq:source_coding_rate_distortion_function} for some distortion level $D(\beta)$.
With an initial guess for $\theta_0$, we can directly apply gradient descent, together with the reparameterisation trick, to optimise \Cref{eq:source_coding_rate_distortion_objective}, using Monte Carlo estimates of expectations where appropriate. 
Here, we ignore some practical details, such as computing the coding distribution $P_{\hat{X}}$ and the mutual information term in \Cref{eq:source_coding_rate_distortion_objective}; for these, we refer the reader to the excellent monograph \cite{yang2023introduction}.
\par
We cannot directly optimise \Cref{eq:source_coding_rate_distortion_objective} when quantisation is our lossy mechanism, because it does not admit a differentiable reparameterisation.
Hence, contemporary learned compression methods that do not use relative entropy coding must resort to approximations.
The two most popular approaches are 1) to perform straight-through estimation (STE) for gradient descent, and 2) to use an additive noise approximation.
STE-based approaches make the ``approximation'' during training that the gradient of the quantiser $\mathcal{Q}$ is the identity instead of being zero almost everywhere: they ``set'' $\nabla_x \mathcal{Q}(x) = 1$ during backpropagation.
Additive noise-based approaches resort to scalar quantisation and, during training only, approximate $\round{x} \approx x + U$, where $U \sim \Unif(-1/2, 1/2)$.
Once trained, they switch to hard quantisation for compression.
\subsubsection{Case Study: Implicit Neural Representations (INRs)}
\par
The motivation for INRs is simple: virtually all data can be thought of as a function mapping some coordinates to signal values.
For example, we may conceive of colour images as functions taking $(x, y)$ pixel coordinates to $(r, g, b)$ colour intensity triplets.
An INR approximates this function using a small neural network $g(x, y \mid \theta)$, effectively ``memorising'' the data coordinate-by-coordinate.
\par
INRs give rise to a compelling approach to data compression: first fit a model to data $X$, then encode the model parameters $\theta$. Direct quantisation, however, poses a challenge, since running gradient descent on quantised parameters is not straightforward. This motivates the use of relative entropy coding. Rather than learning deterministic parameters, we learn a parametric posterior distribution $P_{\theta \mid X}$. 
For example, the typical choice is the humble Gaussian: $P_{\theta \mid X} = \Normal(\mu, \Sigma)$.
We use \Cref{eq:source_coding_rate_distortion_objective} as the loss function to learn the distribution parameters, such as $\mu$ and $\Sigma$ in the Gaussian case, via gradient descent and the reparameterisation trick.
Finally, we encode a sample from $P_{\theta \mid X}$ using relative entropy coding \cite{guo2023compression}.%
\par
Note that Bayesian INRs are an unusual form of transform coding: the representation of $X$ is a draw $\theta$ from the neural network weight distribution, which we learn with gradient descent and encode using relative entropy coding!
\subsubsection{Case Study: Model Compression for Communication-Efficient Federated Learning}
Federated learning enables collaborative model training without centralising clients' raw data, but suffers from high communication costs as clients repeatedly transmit model updates to a central server.
A particularly elegant application of relative entropy coding arises in \textit{federated probabilistic mask training} (FedPM)~\cite{isik2024adaptive}. 
Rather than training all network parameters, clients identify a high-performing \textit{subnetwork} within a shared, randomly initialised network.
The random initialisation is generated from a common seed known to both server and clients. At each iteration, each client trains a local probability mask specifying which parameters to retain, then communicates a sample from this mask distribution to the server, rather than sending the probabilities themselves or the locally sampled mask. Using relative entropy coding, the communication cost decreases from approximately one bit per parameter (for sending a locally sampled mask) to the relative entropy between the client's local mask distribution and the server's global mask distribution. More strikingly, as training progresses, local and global distributions converge, progressively reducing communication costs.
Experiments demonstrate up to 82-fold bitrate reductions compared to direct mask transmission, corresponding to over 2,600-fold overall compression~\cite{isik2024adaptive}.%
\subsection{Compression with realism constraints}
\label{sec:realism_constraints}
\par
In lossy image or video compression, in addition to reconstructing the semantic details of the original data, we also want it to \textit{look good} at all bit rates.
At high rates, this is usually not an issue, but what about low rates?
The key insight is that high distortion does not preclude the reconstruction from looking good.
A landmark insight came from \cite{blau2019rethinking}, which introduced the notion of \textit{realism}.
A compressor's realism depends solely on its output distribution $P_{\hat{X}}$, and we call any sample from the data distribution $P_X$ perfectly realistic.
However, requiring that our compressor be perfectly realistic is generally unrealistic.
Thus, to relax perfect realism, we quantify it with a statistical distance $\Delta(P_X, P_{\hat{X}})$, and extend the rate-distortion function to include this constraint:
\begin{align*}
R(D, \delta) &= \inf_{P_{\hat{X} \mid X} \in \mathcal{D}} \MI{X}{\hat{X}} \\ 
&\text{s.t.}\,\,\, \Exp_{X}[\Exp_{\hat{X} \mid X}[d(X, \hat{X})]] \leq D \text{ and } \Delta(P_{X}, P_{\hat{X}}) \leq \delta
\end{align*}
Here, relative entropy coding provides another benefit over quantisation: there are cases where $R(D, \delta)$ is \textit{strictly} lower if we use relative entropy coding than if we quantise.
We refer readers interested in the theory of realism to Hamdi and G{\"u}nd{\"u}z's article \cite{hamdi2026survey} appearing in the same issue.
\subsection{Variable-rate data compression with diffusion models}
\label{sec:diffc}
\par
\added{%
In this section, we discuss one of the most exciting practical applications of relative entropy coding, namely Theis et al.'s compression with Gaussian diffusion models (DiffC, \cite{theis2022lossy}).
Diffusion models have become a mainstream generative model in recent years, and here we shall review only the bare necessities; we direct readers interested in more details to the recent monograph \cite{lai2025principles}.
}%
\par
\added{%
Given some source $X \sim P_X$ over $\Reals^n$, Gaussian diffusion models consider a \textit{family} of perturbations of $X$ indexed by $t \in [0, 1]$, using a Wiener process $\epsilon_t$, and a function $\sigma_t$ that increases in $t$ with $\sigma_0 = 0$ and $\sigma_1 = 1$, called the \textit{noise schedule}:%
}%
\footnote{It is more common in practice to use the more general formula $Y_t = \alpha_t X + \beta_t \epsilon_t$, where the \textit{interpolants} $\alpha_t$ and $\beta_t$ are positive functions subject to mild conditions, but here we use $\sigma_t$ for simplicity. 
}%
\begin{align}
\label{eq:perturbed_data}
Y_t \defeq \sqrt{1 - \sigma_t^2}\, X + \sigma_t \epsilon_t
\end{align}%
\added{%
Given only the perturbed data $Y_t$, what is the best guess for what $X$ might have been? 
If our discrepancy measure is mean squared error (MSE), then the optimal estimator is the conditional expectation $\Exp[X \mid Y_t]$. 
Thus, diffusion models learn some parametric \textit{denoiser} $f_\theta$ so that ${f_\theta(y\mid 0, t) \approx \Exp[X \mid Y_t = y]}$.
Due to the special nature of Gaussian diffusion, for any ${0 \leq s < t \leq 1}$ we can re-use the denoiser $f_\theta(y \mid 0, t)$ to construct an estimator ${f_\theta(y \mid s, t) \approx \Exp[Y_s \mid Y_t = y]}$.%
}%
\par
\added{%
Diffusion models also make the denoiser probabilistic.
Namely, for small $\Delta t > 0$, we may assume the \textit{denoising error} is Gaussian (this can be justified \cite{lai2025principles})
\begin{align}
\hat{\epsilon}_t = (Y_t - \Exp[Y_t \mid Y_{t + \Delta t}]) \approx \Normal\big(0,\, (\hat{\sigma}_t^{t+\Delta t})^2 \cdot I\big)
\end{align}
for some appropriate standard deviation $\hat{\sigma}_t^{t+\Delta t}$.
For ${s < t}$, we use this approximation to define the \textit{denoising process} 
\begin{align}
\label{eq:emp_denoising_process}
Z_s^t(y) \defeq f_\theta(y \mid s, t) + \hat{\sigma}_{s}^t \cdot \epsilon_{s}^t, \quad \epsilon_{s}^{t} \sim \Normal(0, I)
\end{align}%
However, if we aim to denoise, why would we ever add noise back into the process, as \Cref{eq:emp_denoising_process} does?
The answer is that it allows us to draw samples.
Concretely, fix a subdivision of $[0, t]$ into a mesh $0 = t_0 < t_1 < \cdots < t_n = t$.
Then, define the \textit{ancestral sampler}
\begin{align}
\label{eq:diffusion_ancestral_sampling}
Z^t(z) \,\defeq\, (Z_{t_0}^{t_1} \circ Z_{t_1}^{t_2} \circ \dotsc \circ Z_{t_{n - 1}}^{t_n})(z)
\end{align}
where $f \circ g$ denotes the composition of functions $f$ and $g$, that is, $(f \circ g)(z) \defeq f(g(z))$.
Then, it can be shown that if the denoiser is perfect, that is ${f_\theta(y\mid s, t) = \Exp[Y_s \mid Y_t = y]}$ and the mesh becomes infinitely fine, then for $\epsilon_1 \sim \Normal(0, I)$ we have $Z^1(\epsilon_1) \sim P_X$, meaning that we recover the true data distribution \cite{lai2025principles}.%
}%
\par
\added{%
Diffusion models provide a natural way to construct variable-rate data compression schemes: for all $t \in [0, 1]$, we can treat $Y_t \mid X$ as a lossy version of $X$, and which costs roughly $\MI{X}{Y_t}$ bits to encode.
Specifically, $t=0$ corresponds to lossless data compression, and $t = 1$ to deleting all information about $X$.
Indeed, we can select any level $t$, and use the denoiser $f_\theta(Y_t \mid 0, t)$ to reconstruct $X$.
However, instead of using a deterministic denoiser, we can also use the ancestral sampling method in \Cref{eq:diffusion_ancestral_sampling} for reconstruction, to compute $Z^t(Y_t)$.
As we discussed in the paragraph above, for a sufficiently fine mesh we get a very good approximation $Z^t(Y_t) \approx P_X$, meaning that the Gaussian diffusion-based sampler is essentially perfectly realistic (\Cref{sec:realism_constraints}) at the cost of some extra compute!%
}%
\par
\added{%
However, the benefits of diffusion models do not stop here: they also offer a natural way to perform progressive compression.
Namely, for a mesh $s = t_0 < \dotsc<t_n=t$, let $Y_{s:t} = (Y_{t_0}, Y_{t_1}, \cdots, Y_{t_n})$ and $Z_{s:t} = (Z_{t_0}, \dotsc, Z_{t_n})$.
Then, when the denoiser is perfect, from \Cref{eq:diffusion_ancestral_sampling} we can derive the following identity:
\begin{align}
\KLD{P_{Y_s}}{P_{Z_s}} = \KLD{P_{Y_{s:t}}}{P_{Z_{s:t}}}
\end{align}
Of course, our denoiser will never be perfect in practice, but we can still expect the two sides of the above equation to be close.
Hence, for any common discretisation of the paths, we can replace a single REC problem of large $\KLD{P_{Y_s}}{P_{Z_s}}$, for which exact coding would be infeasible due to large runtime, by a chain of $n$ smaller REC problems along the trajectory. The per-step KL divergences still sum to the original total (by the chain rule), but each is individually smaller, making per-step coding feasible.
}%
\par
\added{%
While DiffC has great potential, it faces several practical implementation challenges, though there has been steady progress in tackling these.
First, since $Y_s$ and $Z_s$ are both Gaussian with different means but identical variances at each step, their density ratio is unbounded.
Hence, in practice, an approximate sampler must be used, such as a step-limited selection sampler (\Cref{sec:approximate_stochastic_codes}).
As we mentioned before, step-limited samplers have the additional benefit of making the encoding process much more parallelisable.
This parallelisability was exploited in \cite{vonderfecht2025lossy}, which implemented a sampling algorithm using custom code for a parallel computing environment (specifically, a CUDA kernel on an nVidia GPU) and reduced the wall-clock encoding time for medium-resolution images from hours to a few seconds.  
Other works have also experimented with more esoteric, less well-principled methods, such as diffusion models with uniform noise \cite{yang2025progressive}, with mixed success.
}%
\subsection{Compressing differential privacy mechanisms}
A pressing concern in data science and machine learning is protecting users' data privacy.
\textit{Local differential privacy} provides a formal framework for this: each user independently perturbs their own data $X$ to produce a noisy version $Y$ before transmission, ensuring that the recipient learns little about the true value.
Formally, a mechanism $M$ mapping $X$ to $Y$ is $\varepsilon$-locally differentially private if for all pairs of possible inputs $x, x'$ and all events $S$:
\begin{align*}
\Prob[M(x) \in S] \leq e^\varepsilon \cdot \Prob[M(x') \in S]
\end{align*}
This guarantees plausible deniability: an adversary observing $Y$ cannot confidently distinguish the true input value.
\par
A prominent application is \textit{federated learning}, in which a central server trains a model using distributed gradient descent.
Clients compute gradient updates from local data and must transmit them to the server without revealing sensitive information.
Popular mechanisms include the additive Gaussian mechanism, $Y = X + \epsilon$ with $\epsilon \sim \Normal(0, \sigma^2)$. 
\par
Since users send a noisy version $Y$ of their data $X$, this provides another natural application for relative entropy coding: we can express the privacy mechanism as $P_{Y \mid X}$, and encode efficiently. However, we cannot naively apply any relative entropy coding algorithm to the privacy mechanism; we must ensure that the code (not just the sample it encodes) does not reveal too much about the user's data.
This issue was recently solved by \cite{liu2024universal}, which developed the first privatised relative entropy coding algorithm capable of encoding exact samples from any privacy mechanism.

\begin{figure}[t]
\centering
\begin{tikzpicture}[auto, node distance=3.2cm, thick, 
  main node/.style={circle, draw, line width=0.3mm, font=\sffamily\small, minimum size=1.6cm, inner sep=0mm}, 
  env node/.style={rectangle, rounded corners=2mm, draw, line width=0.3mm, font=\sffamily\small, minimum size=1.4cm}] 

  \node[env node] (env) {\begin{tabular}{c}Environment\end{tabular}};
  \node[main node] (ctrl) [below left of=env, node distance=2.8cm] {\begin{tabular}{c}\large\faUser\\[-2pt]\scriptsize\faGamepad\end{tabular}};
  \node[main node] (actor) [below right of=env, node distance=2.8cm] {\large\faRobot};

  \draw[->, line width=0.3mm] 
    (env) to[bend right=18] 
    node[midway, left, rotate=45, anchor=south, font=\sffamily\footnotesize] {\begin{tabular}{c}state,\\reward\end{tabular}} 
    (ctrl);

  \draw[->, line width=0.3mm] 
    (env) to[bend right=18] 
    node[midway, right, rotate=-45, anchor=north, font=\sffamily\footnotesize] {state} 
    (actor);

  \draw[->, line width=0.3mm, dashed] 
    (ctrl) -- (actor) 
    node[midway, below, font=\sffamily\footnotesize] {message};

  \draw[->, line width=0.3mm] 
    (actor) to[bend right=18] 
    node[midway, right, rotate=-45, anchor=south, font=\sffamily\footnotesize] {action} 
    (env);

  \node[below=0.15cm of ctrl, font=\sffamily\footnotesize] {Controller};
  \node[below=0.15cm of actor, font=\sffamily\footnotesize] {Actor};

\end{tikzpicture}
\caption{Remote reinforcement learning: a controller (with reward access) guides a remote actor over a rate-constrained channel.}
\label{fig:remote_rl}
\end{figure}
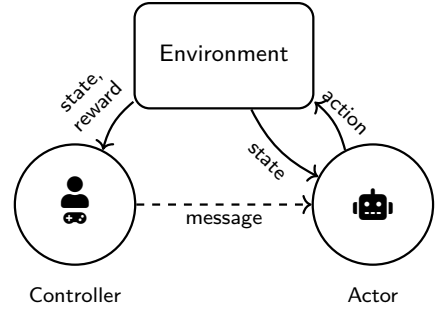
\subsection{Communication-efficient reinforcement learning}
In many reinforcement learning scenarios, the agent executing actions lacks direct access to the reward signal (see \Cref{fig:remote_rl} for an illustration).
This situation arises in human-in-the-loop systems where rewards require human evaluation, in multi-agent settings where rewards depend on collective performance unavailable to individual actors, or when reward computation requires resources (such as large vision-language models) that edge devices lack. These scenarios motivate \textit{remote reinforcement learning}~\cite{kobus2025remote}: a controller observes states and rewards, learns an optimal policy, and must guide remote actors that observe only the state to take appropriate actions over a communication-constrained channel (e.g., a wireless link).
\par
A naive approach is to transmit either the reward or the desired action directly, but both are generally real-valued, which requires costly quantisation.
Relative entropy coding offers a more efficient alternative.
The key insight is that the actor does not need a \textit{specific} action, but rather \textit{any sample} from the controller's policy.
Using relative entropy coding, the actor generates candidate actions from its own policy, and the controller transmits only a short index identifying which candidate to select, thereby enabling exact sampling from the controller's policy. Crucially, the actor simultaneously learns the controller's policy via behavioural cloning from the communicated actions.
As the actor's policy converges to the controller's, the two distributions become more aligned, progressively reducing communication costs.
Experiments demonstrate 12-fold reductions compared to transmitting actions directly and 41-fold reductions compared to transmitting rewards.
\section{Limitations and Future Directions}
This article provided a brief overview of the theory and applications of stochastic codes and relative entropy coding.
However, if relative entropy coding really is the cat's pyjamas, with so many advantages over quantisation, why is it not used in practice? This question is somewhat unfair, as its applicability to practical data compression was only recognised a few years ago.
Nonetheless, relative entropy coding currently possesses two significant disadvantages: speed and synchronisation.
\par
All currently known relative entropy coding algorithms applicable to practical problems are either too slow or too limited, making the design of fast, special-purpose relative entropy codes a major open problem.
\par
The second issue is engineering complexity: REC relies on shared randomness, typically implemented with a shared PRNG seed, which requires the encoder and decoder states to remain perfectly synchronised; desynchronisation leads to catastrophic decoding errors.
While this is ``only'' a matter of rigorous bookkeeping, it remains a monumental underinvestigated engineering challenge.
\par
\added{%
Finally, looking ahead, it is interesting to consider which other areas might be affected by relative entropy coding beyond data compression/information theory.
As we discussed in \Cref{sec:approximate_stochastic_codes}, relative entropy codes can help uncover new results and algorithms in computational statistics more broadly.
Another potential application area is speculative decoding, which has recently been linked to relative entropy codes and Tunstall codes in \cite{kobus2025speculative}.
}%
\section{Acknowledgements}
The authors acknowledge financial support from Imperial College London through an Imperial College Research Fellowship grant awarded to GF, and from the UKRI for projects AI-R (ERC Consolidator Grant, EP/X030806/1) and INFORMED-AI (EP/Y028732/1). We thank Jiajun He for his helpful comments, which helped improve \Cref{sec:diffc}. 
%
%
%
\bibliographystyle{ieeetr}
\bibliography{references}
\end{document}

%% file: math_commands.tex
\usepackage{amsmath,amsfonts,bm}

\def\eqref#1{equation~\ref{#1}}

\def\floor#1{\lfloor #1 \rfloor}
\def\1{\bm{1}}

\DeclareMathAlphabet{\mathsfit}{\encodingdefault}{\sfdefault}{m}{sl}
\SetMathAlphabet{\mathsfit}{bold}{\encodingdefault}{\sfdefault}{bx}{n}

\newcommand{\KL}{D_{\mathrm{KL}}}

\DeclareMathOperator*{\argmin}{arg\,min}

%% file: macros.tex
\usepackage{amsthm}
\usepackage{mathtools}

\DeclarePairedDelimiterX{\infdivx}[2]{[}{]}{%
  #1\;\delimsize\|\;#2%
}
\DeclarePairedDelimiterX{\infdivmi}[2]{[}{]}{%
  #1\;;\;#2%
}
\newcommand{\KLD}{\KL\infdivx} 
\newcommand{\MI}{I\infdivmi} 
\newcommand{\TV}{D_{\mathrm{TV}}}
\newcommand{\TVD}{\TV\infdivx}

\DeclarePairedDelimiter{\norm}{\lVert}{\rVert}
\DeclarePairedDelimiter{\abs}{\lvert}{\rvert}
\DeclarePairedDelimiterX{\innerProd}[2]{\langle}{\rangle}{%
    #1,#2%
}

\def\Normal{\mathcal{N}}
\def\Oh{\mathcal{O}}

\newcommand{\Reals}{\mathbb{R}}

\newcommand{\Nats}{\mathbb{N}}
\newcommand{\Ints}{\mathbb{Z}}

\newcommand{\Prob}{\mathbb{P}}

\newcommand{\Ind}{\mathbf{1}}
\newcommand{\Unif}{\mathrm{Unif}}
\newcommand{\Exp}{\mathbb{E}}

\newcommand{\XSpace}{\mathcal{X}}
\newcommand{\YSpace}{\mathcal{Y}}

\newcommand{\Geom}{\mathrm{Geom}}
\newcommand{\Exponential}{\mathrm{Exp}}

\DeclareMathOperator{\lb}{\mathrm{lb}}
\DeclarePairedDelimiter{\round}{\lfloor}{\rceil}
\DeclareMathOperator{\enc}{\mathtt{enc}}
\DeclareMathOperator{\dec}{\mathtt{dec}}
\DeclareMathOperator{\indep}{\perp}
\DeclareMathOperator{\defeq}{\triangleq}

\newcommand{\RS}{\mathrm{RS}}
\newcommand{\AS}{\mathrm{AS}}
\renewcommand{\SS}{\mathrm{SS}}
\newcommand{\DQ}{\mathrm{DQ}}
\newcommand{\LQ}{\mathrm{LQ}}
\newcommand{\Ent}{\mathbb{H}}
\newcommand{\sort}{\mathtt{sort}}
\newcommand{\Loss}{\mathcal{L}}
\newcommand{\DistFamily}{\mathcal{F}}

\newcommand{\quantiser}{\mathcal{Q}}